\documentclass[sigconf]{acmart}
\usepackage{multirow}
\usepackage{tabularx}
\usepackage{booktabs}
\usepackage[linesnumbered,ruled]{algorithm2e}
\usepackage{tabularx}

\AtBeginDocument{%
  }

\copyrightyear{2025}
\acmYear{2025}
\setcopyright{cc}
\setcctype{by}
\acmConference[MM '25]{Proceedings of the 33rd ACM International Conference on Multimedia}{October 27--31, 2025}{Dublin, Ireland}
\acmBooktitle{Proceedings of the 33rd ACM International Conference on Multimedia (MM '25), October 27--31, 2025, Dublin,
Ireland}\acmDOI{10.1145/3746027.3758164}
\acmISBN{979-8-4007-2035-2/2025/10}

\begin{document}

\title{One Size Fits All? A Modular Adaptive Sanitization Kit (MASK) for Customizable Privacy-Preserving Phone Scam Detection}

\author{Kangzhong Wang}
\affiliation{%
  \institution{The Hong Kong Polytechnic University}
  \city{Hong Kong}
  \country{China}}
\email{kangzhong.wang@connect.polyu.hk}

\author{Zitong Shen}
\affiliation{%
  \institution{The Hong Kong Polytechnic University}
  \city{Hong Kong}
  \country{China}}
\email{esther.shen@connect.polyu.hk}

\author{Youqian Zhang}
\authornote{Corresponding author.}
\affiliation{%
  \institution{The Hong Kong Polytechnic University}
  \city{Hong Kong}
  \country{China}}
\email{you-qian.zhang@polyu.edu.hk}

\author{Michael MK Cheung}
\affiliation{%
  \institution{The Hong Kong Polytechnic University}
  \city{Hong Kong}
  \country{China}}
\email{man-ki-michael-mk.cheung@polyu.edu.hk}

\author{Xiapu Luo}
\affiliation{%
  \institution{The Hong Kong Polytechnic University}
  \city{Hong Kong}
  \country{China}}
\email{csxluo@comp.polyu.edu.hk}

\author{Grace Ngai}
\affiliation{%
  \institution{The Hong Kong Polytechnic University}
  \city{Hong Kong}
  \country{China}}
\email{grace.ngai@polyu.edu.hk}

\author{Eugene Yujun Fu}
\affiliation{%
  \institution{The Education University of Hong Kong}
  \city{Hong Kong}
  \country{China}}
\email{eugenefu@eduhk.hk}

\renewcommand{\shortauthors}{Kangzhong Wang et al.}

\begin{abstract}
Phone scams remain a pervasive threat to both personal safety and financial security worldwide. 
Recent advances in large language models (LLMs) have demonstrated strong potential in detecting fraudulent behavior by analyzing transcribed phone conversations. 
However, these capabilities introduce notable privacy risks, as such conversations frequently contain sensitive personal information that may be exposed to third-party service providers during processing.
In this work, we explore how to harness LLMs for phone scam detection while preserving user privacy. 
We propose MASK (Modular Adaptive Sanitization Kit), a trainable and extensible framework that enables dynamic privacy adjustment based on individual preferences. 
MASK provides a pluggable architecture that accommodates diverse sanitization methods—from traditional keyword-based techniques for high-privacy users to sophisticated neural approaches for those prioritizing accuracy. 
We also discuss potential modeling approaches and loss function designs for future development, enabling the creation of truly personalized, privacy-aware LLM-based detection systems that balance user trust and detection effectiveness, even beyond phone scam context.

\end{abstract}




\maketitle

\section{Introduction}

Phone scams remain a persistent and evolving threat to individuals and society. 
By impersonating authorities, family members, or service providers, scammers deceive victims over the phone for illegal financial gain. 
This form of fraud continues to cause significant economic damage: for instance, recent reports estimate that in 2024 alone, phone scams led to trillions of dollars in losses globally~\cite{rogers2024international}. 
Moreover, victims may also suffer psychological trauma, anxiety, and long-term distrust in communication channels~\cite{buse2023unveiling, hu2022btg, hu2024gat}.

Efforts to prevent phone scams have been ongoing for years. 
Governments and organizations have adopted measures such as public awareness campaigns~\cite{burke2022educational, deliema2020financial, smith2008raising,jensen2024awareness} and technical solutions like caller ID verification~\cite{truecaller2024}. 
However, as defensive mechanisms improve, so too do the tactics of scammers, who constantly find ways to bypass existing safeguards~\cite{mustafa2018end, tu2019users}.
In recent years, researchers have begun to shift their focus toward the content of scam communications~\cite{peng2018fraud,tseng2015fraudetector, chadalavada2024distinguishingscamsfraudensemble, chang2024exposingllmvulnerabilitiesadversarial, shen2024combatingphonescamsllmbased, shen2025itwarnedjustright}. 
Since scammers must engage their targets in conversation, analyzing these interactions has emerged as a promising detection strategy. 
With the rapid advancement of large language models (LLMs), it has become increasingly feasible to analyze voice-to-text transcriptions of phone calls and detect scam patterns with high accuracy. This approach extends to processing multimedia content, leveraging the rich information embedded in audio and potentially other modalities. Several recent studies~\cite{shen2024combatingphonescamsllmbased, shen2025itwarnedjustright, ma2025teleantifraud, chang2024exposing} have successfully demonstrated the effectiveness of LLMs in identifying scam-related content in conversations (see more details in Section~\ref{sec:background_llm-based_phone_scam_detection}).

\begin{figure}[t]
    \centering
    \includegraphics[width=1\linewidth]{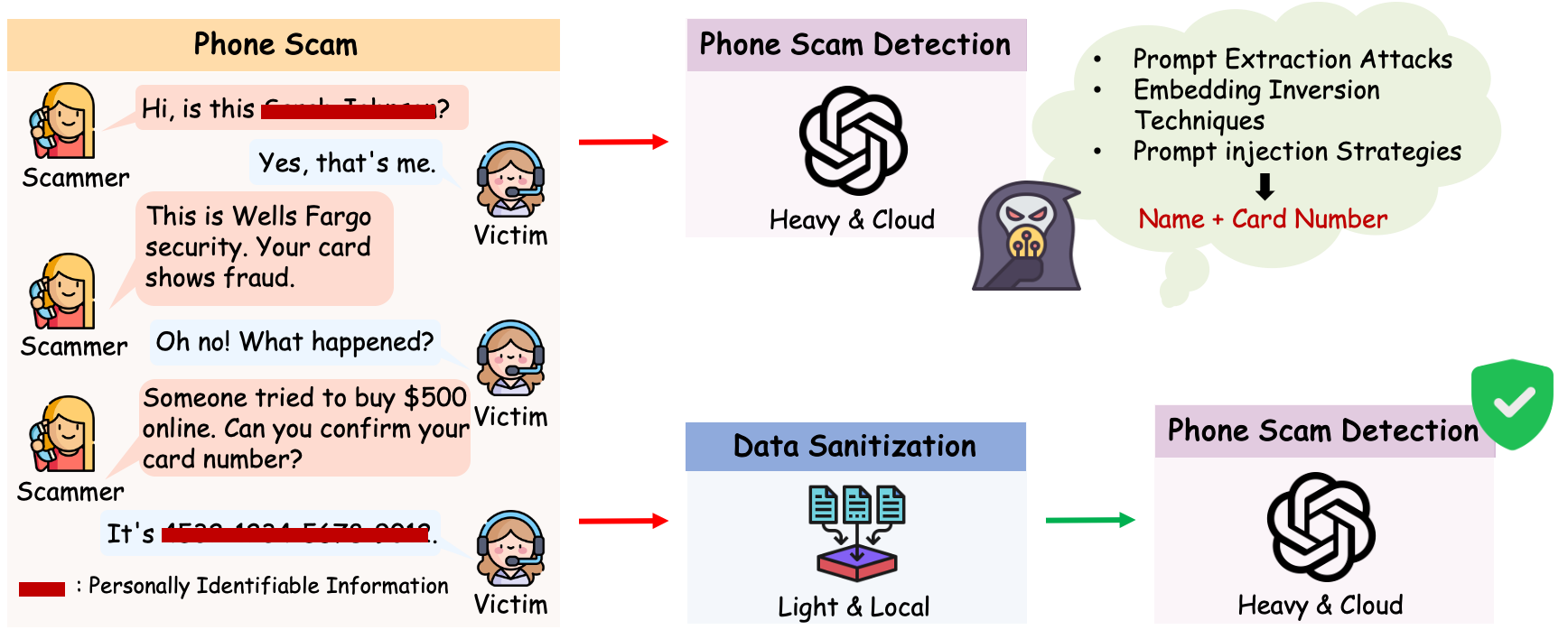}
    \caption{LLM-based phone scam detection approaches. Cloud-based models possibly risk privacy leakage, while data sanitization (at user/local devices) enables safe detection with cloud models.}
    \label{fig:phone_scam_detection_comparison}
\end{figure}

However, conversations between a caller and recipient often contain sensitive information, and the LLM-based approach introduces a serious concern, user privacy. 
As shown in Fig. ~\ref{fig:phone_scam_detection_comparison}, sending these conversations to a centralized LLM (e.g., a cloud-based server) for analysis can expose users to significant privacy risks: If such data were to be leaked or misused, the consequences could be severe, ranging from identity theft to further exploitation (more discussion in Section~\ref{sec:privacy_challenges_in_cloud-based_llms}).
It's crucial to note that traditional encryption does not fully solve the privacy problem, as encrypted data still needs to be decrypted before being processed by an LLM, during which the sensitive content is exposed.

%On the other hand, as just mentioned, most current LLM deployments are still cloud-based. 
To mitigate privacy risks in LLM-powered applications, recent work has proposed prompt sanitization techniques that remove or mask sensitive information before sending it to the model~\cite{carvalho2023tem, arnold2023driving, arnold2023guiding, chen2023customized, utpala2023locally, casper2024promptsanitization}. 
However, the context of phone scam detection introduces unique challenges. 
Conversations in this setting are often highly sensitive, involving personal, financial, or emotional information, and the sensitivity of the content evolves dynamically throughout the conversation process. 
Moreover, real-time scam detection requires low-latency processing and minimal user intervention, making traditional interactive sanitization interfaces (e.g., highlight-and-adjust UIs) impractical.
Further, users’ expectations and preferences for privacy vary widely, influenced by their personal values, contexts, and risk perceptions, and could be shaped in real-time as the conversation unfolds.
Prior research has shown that users’ privacy needs in LLM-based applications are deeply subjective and context-dependent~\cite{zhang2024s, li2024human, zhou2025rescriber}.

We argue that protecting privacy in phone scam detection should \textit{center user autonomy and preference} and \textit{go beyond static sanitization rules}, which most follow a ``one-size-fits-all'' paradigm, failing to account for the subjective, contextual, and evolving nature of privacy preferences. 
We propose the Modular Adaptive Sanitization Kit (MASK), a user-centric, configurable privacy framework designed to support diverse privacy needs in real-time. 
MASK marks a brave and novel departure from conventional thinking: Rather than enforcing a fixed sanitization pipeline, MASK embraces privacy pluralism by allowing users to dynamically configure how much and what kind of information gets sanitized across multimedia content, such as voice and text data.
It respects the user's autonomy, adapts to their evolving context, and supports interactive, real-time privacy decisions, all while maintaining the utility needed for accurate phone scam detection.

\section{Background}
\label{sec:background}

% In this section, we present background on using LLM to detect phone scams, privacy challenges in cloud-based LLM, and data sanitization. 
%Telecommunications fraud continues to plague individuals and financial institutions, causing devastating financial losses and psychological harm. Traditional approaches to phone scam detection, often relying on rule-based systems or statistical models, have struggled to keep pace with the evolving sophistication of scam tactics. The recent advancements in large language models (LLMs) have opened new avenues for combating these threats, particularly through their capacity for nuanced semantic analysis of conversational data. However, the deployment of LLMs for real-time or post-call phone scam detection introduces a critical tension between efficacy and privacy.

\subsection{LLM-Based Phone Scam Detection}
\label{sec:background_llm-based_phone_scam_detection}

A few studies have explored the use of large language models (LLMs) for real-time detection of telecom fraud through the analysis of phone call transcripts. 
Shen et al.~\cite{shen2025itwarnedjustright} proposed a framework that evaluates fraudulent intent in conversations and delivers immediate warnings to users. 
Their work highlights a key trade-off between detection accuracy and response timeliness but remains focused solely on textual input.
In another work, Shen et al.~\cite{shen2024combatingphonescamsllmbased} examined the strengths and limitations of LLMs in this domain, identifying challenges such as data bias, hallucinations, and low recall, and emphasizing the need for high-quality, diverse training datasets. 
In a follow-up work, Ma and Wang et et al.,~\cite{ma2025teleantifraud} create a synthetic phone scam datasets to facilitate extensive studies in this area.
Chang et al.,~\cite{chang2024exposing} further investigated the vulnerability of LLMs to adversarial scam messages, constructing a dataset of original and manipulated texts to analyze misclassification rates.

%The inherently conversational nature of phone scams makes them particularly amenable to analysis by large language models (LLMs). Recent research has demonstrated the promising capabilities of LLMs in identifying fraudulent intent and detecting scam patterns by analyzing transcribed phone conversations \cite{shen2024combatingphonescamsllmbased, shen2025itwarnedjustright}. These models can discern subtle linguistic cues, emotional shifts, and narrative inconsistencies that are indicative of fraudulent activity, far surpassing the capabilities of conventional methods. This superior performance is attributed to their training on diverse, large-scale corpora, enabling them to capture complex linguistic patterns that signal deceit \cite{brown2020languagemodels}.

\subsection{Privacy Challenges in Cloud-Based LLMs}
\label{sec:privacy_challenges_in_cloud-based_llms}
%In a phone call, we use natural language to onstruct identities and communicate our information, where is very likely to include much personal privacy information. When we use llms to detect phone scam by accessing use's phone conversation directly, it's important that llms do not violate human notions of privacy or make use of data in ways beyond what is needed for the utility of the technology\cite{nissenbaum2009privacy}\cite{zuboff2019age}. 
Deploying large language models (LLMs) in cloud environments raises significant privacy concerns, particularly when users submit sensitive data via conversational prompts. These interactions often include personally identifiable information (PII)—such as names, addresses, and financial details—which must be transmitted to LLM service providers for processing~\cite{li2024privacylargelanguagemodels}. However, such data is typically logged and stored by the providers~\cite{li2024privacylargelanguagemodels, zhang2024securityprivacychallenges}, increasing the risk of privacy breaches.
Stored prompts can be exploited through a range of emerging attack vectors, including prompt injection, embedding inversion, and prompt extraction~\cite{carlini2021extractingtrainingdata, liu2023promptinjectionsurvey}. Prompt injection allows attackers to manipulate LLM behavior, potentially eliciting unauthorized disclosure of user inputs. Embedding inversion techniques can reconstruct original prompts from internal model embeddings, while prompt extraction attacks can retrieve user data directly from logged interactions~\cite{carlini2021extractingtrainingdata, zhang2024securityprivacychallenges, liu2023promptinjectionsurvey}.
These threats reveal a critical paradox in phone scam detection: the very AI tools designed to protect users may themselves become sources of privacy risk. This underscores the urgent need for privacy-preserving mechanisms that can mitigate these vulnerabilities while maintaining the effectiveness of LLM-based detection systems.

\subsection{Data Sanitization}

A critical assumption in privacy-preserving LLM applications is that private information can be efficiently identified and removed~\cite{brown2022doesmeanlanguagemodel}. 
Existing sanitization techniques typically rely on rule-based filters, machine learning-based named entity recognition (NER), and local LLMs for topic identification. 
Rule-based methods redact predefined patterns of personally identifiable information (PII), while NER models detect and mask entities such as names and addresses~\cite{casper2024promptsanitization}. 
More recently, lightweight local LLMs deployed on user devices have been used to identify sensitive topics, enabling selective sharing of non-sensitive content with cloud-based services~\cite{casper2024promptsanitization, deprompt2024desensitization}.
Despite their promise, these methods face challenges. 
Over-sanitization can degrade the utility of data for downstream tasks, including scam detection, while under-sanitization risks PII leakage due to the context-dependent subtleties of natural language~\cite{whatdoesitmean2022privacylanguagemodels}. 
Alternative approaches such as differential privacy, attempt to obscure PII by injecting statistical noise. 
However, they often incur high computational costs and struggle to preserve semantic fidelity in nuanced conversational contexts~\cite{whatdoesitmean2022privacylanguagemodels, onprotecting2025privacyleakage}.

\section{Modular Adaptive Sanitization Kit (MASK)} \label{sec:overview_MASK}

\begin{figure*}[t]
    \centering
    \includegraphics[width=0.7\linewidth]{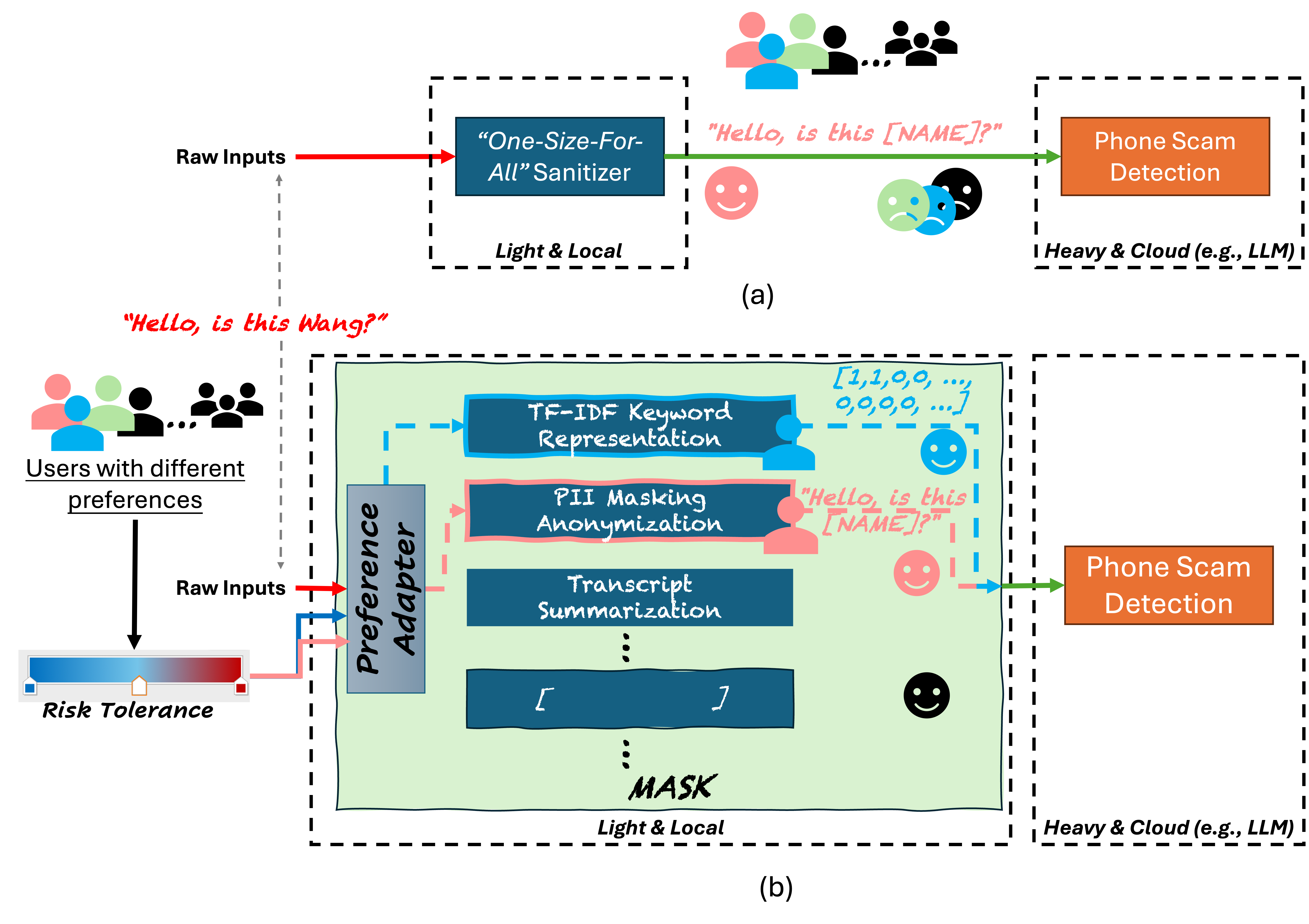}
    \caption{Overview of privacy-preserving frameworks: (a) Existing methods employ fixed (e.g., PII-like entity masking) with one-size-fits-all approaches, offering limited flexibility and no user preference customization. (b) The proposed Modular Adaptive Sanitization Kit (MASK) enables dynamic privacy adjustment through a pluggable architecture that accommodates various sanitization methods such as traditional keyword-based and PII-like methods for users with different privacy preferences. It also provides extensible plugin capabilities for future enhancements, particularly for sophisticated neural network approaches.}
    \label{fig:overview_MASK}
\end{figure*}

To addresses the limitations of existing one-size-fits-all privacy-preserving approaches. This paper proposes and explores the design of the Modular Adaptive Sanitization Kit (MASK). At its core, MASK is a flexible, user-centric privacy-preserving framework that employs a pluggable architecture that allows users to dynamically select and configure sanitization modules based on their individual privacy preferences and contextual requirements. In particular, MASK consists of three primary components (Fig. ~\ref{fig:overview_MASK} (b)): a privacy preference adapter, a modular sanitization layer, and an extensible plugin architecture.

\subsection{Privacy Preference Adapter and Risk Tolerance Parameter}
The privacy preference adapter serves as the central component in MASK that mapping user-defined privacy requirements to modular sanitzation strategies. The adapter supports fine-grained privacy control, particularly, through user-specific risk tolerance.
The "\textit{risk tolerance}" parameter functions analogously to the "\textit{temperature}" setting in commercial LLMs that controls output randomness and creativity. In MASK, \textit{risk tolerance} provides users with intuitive control over sanitization intensity, directly affecting the aggressiveness of privacy-preserving operations.
A low \textit{risk tolerance} (similar to low \textit{temperature}) trigger the MASK to perform over-protection with more aggressive sanitization strategies and techniques, resulting in conservative, high-level of privacy-preserving prompts sent to the cloud server. 

Fig.~\ref{fig:overview_MASK} (b) illustrates this concept using the input phrase "\textit{Hello, is this Wang?}" as an example. Under low risk tolerance settings, the system might apply extreme sanitization by converting the entire input into a tokenized feature vector that only preserves structural information, such as [1, 1, 0, 0, ...] representing the count of different entity types (e.g., one name entity, one question marker, zero location entities, zero phone numbers). This approach maximizes privacy by eliminating all semantic content while retaining minimal structural signals for detection. In contrast, a higher risk tolerance setting would apply selective privacy entity masking, producing outputs like "\textit{Hello, is this [Name]?}" that preserve the conversational structure and context while only masking explicit personal identifiers. We anticipate that users can specify their privacy requirements for MASK through intuitive interfaces that map their  privacy preferences into concrete sanitization behaviors. This parameterization allows the same framework to serve privacy-conscious users who prioritize data protection and accuracy-prioritized users who need maximum detection performance.

Furthermore, we anticipate that the preference adapter can go beyond fixed approaches that rely on static mappings between risk tolerance and specific sanitizers. Instead, it supports more fine-grained privacy control through a multi-dimensional preference model that jointly considers multiple factors such as entity and data sensitivity levels, contextual information, and user risk tolerance in an integrated manner. A particular promising direction is to implement with neural network models that can dynamically balance the trade-off between risk tolerance (privacy preservation rate) and semantic maintenance. This sophisticated balancing can be implemented through joint modeling approaches that optimize both privacy preservation and semantic similarity as components of a unified loss function in future developments.

\subsection{Modular Sanitization Layer Supporting Extensible Plugin}

The modular sanitization layer implements a diverse collection of privacy-preserving methods which organized as interchangeable modules.
Our MASK is initially implemented with four primary sanitization strategies, namely, (1) \textit{TF-IDF Keyword Representation}, (2) \textit{PII Statistical Representation}, (3) \textit{PII Masking Anonymization}, and (4) \textit{Transcript Summarization} (Details in Section~\ref{sec:sanitization_methods}). This offers the initial set of options matching users' different \textit{risk tolerance} levels.
For example, for users demanding maximum privacy protection, the traditional TF-IDF keyword-based and statistical methods that operate without deep semantic understanding, ensuring minimal information leakage at the cost of potential accuracy reduction.

To accommodate future developments and specialized requirements, MASK is implemented with an extensible plugin architecture that allows researchers and developers to integrate new sanitization techniques seamlessly. This forward-looking design includes well-defined APIs and standardized interfaces that enable the incorporation of emerging privacy-preserving methods, from rule-based detection methods to complex neural network approaches. Moreover, the plugin architecture also supports trainable components, allowing sanitization modules to adapt and improve their performance while maintaining user-specified privacy constraints by neural network with multi-task learning.
% \section{Methodology}

\begin{figure}[t]
    \centering
    \includegraphics[width=1\linewidth]{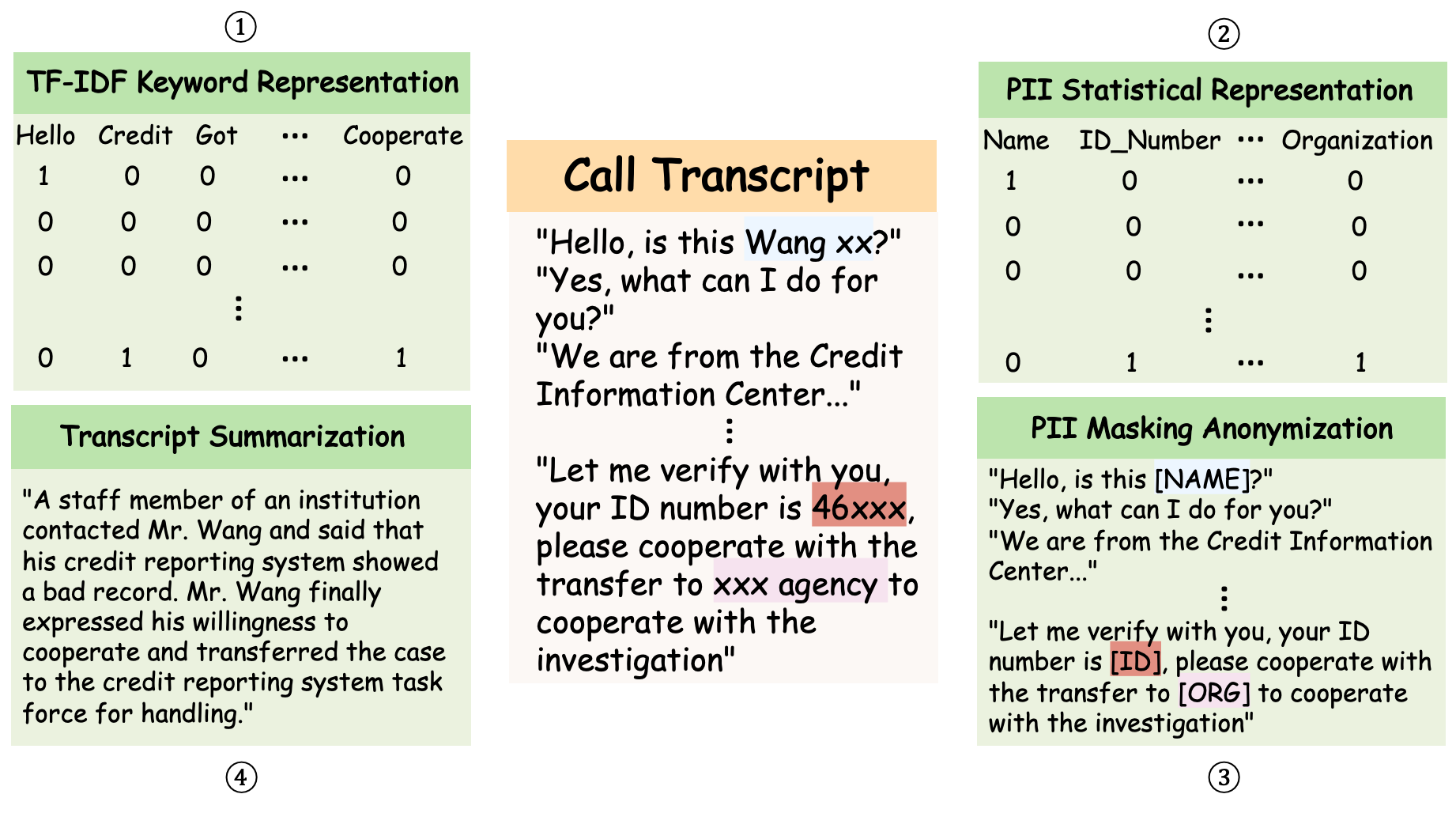}
    \caption{Overview of the four sanitization strategies: (1) TF-IDF keyword representation, (2) PII statistical representation, (3) PII masking anonymization, and (4) transcript summarization.}
    \label{fig:sanitization}
\end{figure}

\section{Privacy-Preserving Sanitization Strategies} 
\label{sec:sanitization_methods}

In this section, we present four primary sanitization strategies that were initially implemented and examined within the Modular Sanitization Layer of MASK. 
It is important to note that these strategies are provided as examples, and the proposed framework offers flexibility for customization, as discussed earlier.
In addition, note that in the following sections of this study, we utilize the datasets provided by~\cite{shen2024combatingphonescamsllmbased}, which consist of data collected from real scam and benign phone calls in Chinese, rather than synthetic data. 
We believe this choice better reflects real-world scenarios and enhances the practical value of this work.

\subsection{TF-IDF Keyword Representation}

Term frequency-inverse document frequency (TF-IDF)~\cite{10.5555/106765.106782} is a widely used statistical method in information retrieval and text mining that quantifies the importance of a word within a document relative to an entire corpus. 
In this study, the so-called ``TF-IDF Keyword Representation'' method converts each conversational transcript into a structured numeric vector format based on the frequency of selected discriminative keywords.

This method highlights words that are more indicative of scam-related content compared to normal conversations, thereby preserving critical semantic information while mitigating the risk of privacy leakage, as shown in Figure~\ref{fig:sanitization}.
Note that to identify the most discriminative features, the difference in mean TF-IDF scores between the two classes (scam v.s. normal phone calls) is calculated, and the top-$N$ keywords with the largest absolute differences are selected. 
For each utterance in a transcript, a fixed-length numeric vector is generated, where each element represents the occurrence frequency of a selected keyword within that utterance.
These numeric vectors serve as privacy-preserving representations for subsequent downstream scam detection tasks using cloud-based LLMs.

% The overall procedure is formally described in Algorithm~\ref{algo:tfidf_representation}.

% \begin{algorithm}[h]
% \caption{TF-IDF Keyword Representation}
% \label{algo:tfidf_representation}
% \KwIn{Transcript $X$, stopword set $S$, punctuation set $P$, pre-selected TF-IDF keyword list $K_{tf-idf}$}
% \KwOut{TF-IDF keyword vector representation $V_{tf-idf}$}

% Initialize empty list $V_{tf-idf}$\\
% Segment transcript $X$ into utterances $U = \{u_1, u_2, \dots, u_n\}$\\
% \ForEach{utterance $u_i$ in $U$}{
%     Tokenize $u_i$ into tokens $T_i$\\
%     Clean tokens: $T'_i \leftarrow \{ t \in T_i \mid t \notin S \cup P, \text{len}(t)>1 \}$\\
%     Initialize keyword-count vector $v_i \leftarrow [0, 0, \ldots, 0]$ with length $|K_{tf-idf}|$\\
%     \ForEach{keyword $k_j$ in $K_{tf-idf}$}{
%         Set $v_{i,j}$ as the frequency of $k_j$ in $T'_i$\\
%     }
%     Append $v_i$ to $V_{tf-idf}$\\
% }
% \Return $V_{tf-idf}$
% \end{algorithm}

\subsection{PII Statistical Representation and Masking Anonymization}
PII refers to personally identifiable information. 
Both PII Statistical Representation and PII Masking Anonymization employ a two-step hybrid approach to in conversational transcripts. As shown in Figure~\ref{fig:sanitization}, each transcript is processed using regular expression-based pattern matching to detect structured PII, such as identification numbers, phone numbers, account identifiers, emails, URLs, bank card numbers, and dates, and a neural named entity recognition (NER) model~\cite{honnibal2017spacy} to capture unstructured PII, including personal names, organizations, and locations. 
All detected entities are replaced with standardized, category-specific placeholders (e.g., \texttt{[ID]}, \texttt{[PHONE]}, \texttt{[NAME]}, \texttt{[ORG]}, \texttt{[LOC]}), with unmatched numeric sequences masked as \texttt{[NUM]}.

The two strategies differ in how they represent the processed transcripts for downstream analysis:
\begin{itemize}
    \item \textbf{PII Statistical Representation} transforms each utterance into a fixed-length vector, where each dimension reflects the count of a specific PII category within the utterance.
    \item \textbf{PII Masking Anonymization} outputs an anonymized transcript in which all identified PII entities are substituted by their category placeholders.
\end{itemize}

By distinguishing between statistical abstraction and structural anonymization, these methods provide flexible options for balancing privacy protection and semantic retention in language-based detection pipelines.

\subsection{Transcript Summarization}

Transcript Summarization is designed to maximize privacy protection and reduce input redundancy by condensing entire conversational transcripts into concise and high-level summaries. These summaries are constructed to retain only the core events and essential information, while systematically omitting extraneous details and all personally sensitive content. This method leverages local large language models (LLMs) to generate privacy-preserving, document-level representations prior to any cloud-based analysis, thereby minimizing the risk of sensitive data exposure. In this study,we select Qwen3-4b~\cite{yang2025qwen3} as our local lightweight LLM for transcript summarization. Qwen3-4b combines a compact model size, which is suitable for efficient deployment on local or resource-constrained devices including smartphones, with robust generation capabilities and high summarization quality on Chinese conversational data. These characteristics make Qwen3-4b particularly well suited for secure, edge-based preprocessing in real-world scenarios. As shown in Figure~\ref{fig:sanitization}, for each transcript, the complete textual content is provided as input to the local LLM. The model is prompted to produce a brief summary that distills the most salient events, actions, and key facts, while explicitly excluding personal names, identifiers, contact information, or subjective commentary. The output is a single, coherent summary statement for each transcript, ensuring that only non-sensitive, task-relevant information is utilized for downstream scam detection analysis.

% Algorithm~\ref{algo:doc_summary} formally describes the procedure for generating a document-level summary from a conversational transcript.

% \begin{algorithm}[h]
% \caption{Transcript Summarization (Document-Level)}
% \label{algo:doc_summary}
% \KwIn{Transcript $T$; local LLM model $M$; summarization prompt template $P$}
% \KwOut{Transcript-level summary $s$}

% Construct privacy-preserving summarization prompt $p$ using template $P$ and transcript $T$\;
% Query local LLM $M$ with prompt $p$ to generate summary $s$\;
% \Return $s$\;
% \end{algorithm}

\subsection{Evaluation Metrics of Sanitization}

%Ideally, user interactions with cloud-based large language models should satisfy two essential requirements. First, the transmission and processing of user data must safeguard sensitive personal information against potential privacy breaches or malicious attacks. Second, privacy preservation must retain sufficient semantic information to ensure that the LLM’s responses faithfully reflect the user’s original intent. Accordingly, in this study, 
We evaluate the effectiveness of our sanitization strategies along two primary dimensions: (1) privacy preservation and (2) semantic retention.

\textbf{\textit{Privacy Preservation.}} To quantitatively assess the privacy-preserving capacity of each sanitization strategy, we introduce the \textit{PII Removal Rate} (PRR). This metric measures the proportion of PII entities that are successfully removed or obfuscated from transcripts by the sanitization process. For each transcript, we employ a comprehensive PII detection pipeline, combining regular expression matching and NER to identify all types of PII in both the original and sanitized versions. The PRR is computed as:
\begin{equation}
    \mathrm{PRR} = 1 - \frac{\sum_{i=1}^{N} |\text{PII}^{(i)}_{\mathrm{sanitized}}|}{\sum_{i=1}^{N} |\text{PII}^{(i)}_{\mathrm{raw}}|}
\end{equation}
where $\text{PII}^{(i)}_{\mathrm{raw}}$ and $\text{PII}^{(i)}_{\mathrm{sanitized}}$ denote the sets of detected PII in the $i$-th transcript before and after sanitization, respectively, and $N$ is the total number of evaluated transcripts. Higher PRR values indicate a greater degree of privacy protection.

\textbf{\textit{Semantic Retention.}} It is crucial that sanitization methods do not excessively degrade the semantic content necessary for downstream LLM-based analysis. To this end, we assess the semantic retention rate (SRR) of each strategy by computing the average semantic similarity between the original and sanitized versions of each transcript:
\begin{equation}
    \mathrm{SRR} = \frac{1}{N} \sum_{i=1}^{N} \mathrm{Sim}(T_i, \hat{T}_i)
\end{equation}
where $T_i$ and $\hat{T}_i$ denote the original and sanitized versions of the $i$-th transcript, respectively, and $\mathrm{Sim}(\cdot, \cdot)$ represents a document-level semantic similarity function. In this study, we employ Sentence-BERT cosine similarity~\cite{reimers2019sentence} to quantify the preservation of semantic content. Higher SR values indicate better retention of core semantic information, thereby allowing the LLM to more accurately interpret user intent and fulfill the intended task.

\section{Experimental Results}
We comprehensively evaluate the effectiveness of the proposed sanitization strategies for phone scam detection on conversational transcripts. Experiments are conducted using four advanced large language models (LLMs) that have demonstrated strong performance in Chinese understanding tasks~\cite{chiang2024chatbot}: ChatGPT-4o (or simply denoted as GPT)\cite{ChatGPT4o}, Gemini-2.5-Flash (or Gemini) \cite{Gemini2.5Flash}, Qwen2.5-MAX (or Qwen) \cite{Qwen2.5MAX}, and DeepSeek-V3 (or DeepSeek)\cite{DeepSeekV3}. To systematically compare the privacy-utility trade-offs of each approach, all sanitization methods are applied to the dataset, and the processed transcripts are used as input for each LLM-based scam detector. Detection performance is reported using standard evaluation metrics, including accuracy (Acc.), precision (P.), recall (R.), and F1 score, across all combinations of data strategy and model. 
The results presented below highlight the comparative robustness of each LLM under the different sanitization methods.

\begin{table}[t]
  \centering
  \setlength{\tabcolsep}{9pt}
  % \small
  \resizebox{\columnwidth}{!}{%
\begin{tabular}{lccccccc}
\hline
\textbf{Sanitization}                   & \textbf{Model} & \textbf{Acc.} & \textbf{P.}   & \textbf{R.}   & \textbf{F1}   & \textbf{PRR}           & \textbf{SRR}           \\ \hline
\multirow{4}{*}{\textbf{Original Text}} & GPT            & 0.97          & \textbf{1.00} & 0.94          & 0.97          & \multirow{4}{*}{0.000} & \multirow{4}{*}{1.000} \\
                                        & Gemini         & 0.96          & 0.96          & 0.96          & \textbf{0.96} &                        &                        \\
                                        & Qwen           & \textbf{0.98} & \textbf{1.00} & \textbf{0.96} & \textbf{0.98} &                        &                        \\
                                        & DeepSeek       & \textbf{0.98} & \textbf{1.00} & \textbf{0.96} & \textbf{0.98} &                        &                        \\ \hline
\multirow{4}{*}{\textbf{\begin{tabular}[c]{@{}l@{}}TF-IDF \\ Keyword \\ Representation\end{tabular}}} &
  GPT &
  \textbf{0.83} &
  0.88 &
  0.76 &
  \textbf{0.82} &
  \multirow{4}{*}{1.000} &
  \multirow{4}{*}{0.168} \\
                                        & Gemini         & 0.49          & 0.49          & \textbf{0.98} & 0.66          &                        &                        \\
                                        & Qwen           & 0.67          & \textbf{1.00} & 0.32          & 0.48          &                        &                        \\
                                        & DeepSeek       & 0.79          & \textbf{1.00} & 0.58          & 0.73          &                        &                        \\ \hline
\multirow{4}{*}{\textbf{\begin{tabular}[c]{@{}l@{}}PII \\ Statistical \\ Representation\end{tabular}}} &
  GPT &
  0.27 &
  0.21 &
  \textbf{0.16} &
  \textbf{0.18} &
  \multirow{4}{*}{1.000} &
  \multirow{4}{*}{0.085} \\
                                        & Gemini         & \textbf{0.53} & \textbf{1.00} & 0.06          & 0.11          &                        &                        \\
                                        & Qwen           & 0.52          & 0.75          & 0.06          & 0.11          &                        &                        \\
                                        & DeepSeek       & \textbf{0.53} & \textbf{1.00} & 0.06          & 0.11          &                        &                        \\ \hline
\multirow{4}{*}{\textbf{\begin{tabular}[c]{@{}l@{}}PII \\ Masking \\ Anonymization\end{tabular}}} &
  GPT &
  0.96 &
  \textbf{1.00} &
  0.92 &
  0.96 &
  \multirow{4}{*}{1.000} &
  \multirow{4}{*}{0.861} \\
                                        & Gemini         & \textbf{0.98} & \textbf{1.00} & 0.96          & \textbf{0.98} &                        &                        \\
                                        & Qwen           & 0.94          & 0.89          & \textbf{1.00} & 0.94          &                        &                        \\
                                        & DeepSeek       & \textbf{0.98} & \textbf{1.00} & 0.96          & \textbf{0.98} &                        &                        \\ \hline
\multirow{4}{*}{\textbf{\begin{tabular}[c]{@{}l@{}}Transcript \\ Summarization\end{tabular}}} &
  GPT &
  0.88 &
  \textbf{1.00} &
  0.76 &
  0.86 &
  \multirow{4}{*}{0.903} &
  \multirow{4}{*}{0.412} \\
                                        & Gemini         & \textbf{0.89} & \textbf{1.00} & \textbf{0.78} & \textbf{0.88} &                        &                        \\
                                        & Qwen           & 0.87          & 0.97          & 0.76          & 0.85          &                        &                        \\
                                        & DeepSeek       & 0.86          & \textbf{1.00} & 0.72          & 0.84          &                        &                        \\ \hline
\end{tabular}%
}
  \caption{Performance of different sanitization strategies and detection models. 
Metrics include accuracy, precision, recall, F1 score, privacy removal rate (PRR), and semantic retention rate (SRR).}
  \label{tab:exp_results_sanitation}
\end{table}

\subsection{Overall Evaluation}

Table~\ref{tab:exp_results_sanitation} presents a systematic comparison of the sanitization strategies in terms of their impact on both privacy protection and detection performance. The evaluation covers all four detection models and captures differences in semantic retention, privacy risk, and the preservation of critical conversational information.

The \textbf{PII Masking Anonymization} approach achieves a strong balance between privacy and detection effectiveness. By removing all identified sensitive information (PRR = 1.000) while largely retaining the semantic content of the original conversation (SRR = 0.861), this method maintains nearly the same F1 scores as the unprocessed baseline. The minimal decline in detection accuracy indicates that replacing explicit entities with category placeholders is sufficient to prevent privacy leakage, while preserving the narrative and interaction cues necessary for reliable classification. This level of performance demonstrates that structured anonymization at the entity level supports both compliance with privacy standards and robust detection results.

By comparison, the \textbf{TF-IDF Keyword Representation} and \textbf{PII Statistical Representation} methods, although highly effective in removing sensitive details (PRR = 1.000), result in considerable degradation of semantic content. With much lower semantic retention rates (SRR = 0.168 and 0.085, respectively) and a pronounced drop in F1 scores, these strategies often discard critical context and cues , making it more difficult for the model to accurately distinguish scams from ordinary conversation. 

The \textbf{Transcript Summarization} approach provides a moderate level of privacy protection (PRR = 0.903) and intermediate semantic retention (SRR = 0.412), with corresponding F1 scores situated between those of the entity masking and abstraction-based methods. Summarization condenses the transcript to its main points, omitting sensitive or redundant material, which can be advantageous for applications requiring smaller inputs or strict privacy requirements. Nevertheless, this reduction in detail sometimes leads to a loss of information relevant for accurate detection, and thus performance does not reach the level observed with entity-level masking.

These results illustrate the fundamental trade-off between privacy protection and semantic utility in the context of transcript sanitization. Entity-level masking stands out as the most effective strategy for balancing both aims, providing strong privacy safeguards with minimal compromise in downstream task performance. Methods based on greater abstraction or summarization, while sometimes necessary for stricter privacy or resource constraints, tend to increase the risk of missing critical context and reducing detection reliability. Careful selection of sanitization techniques is therefore essential, particularly in applications where both privacy and detection accuracy are priorities.

\subsection{Impact of Sanitization on Precision and Recall}

While the accuracy and F1 score offers an overall measure of detection performance, a more detailed analysis of precision and recall provides greater insight into the strengths and weaknesses of each sanitization strategy. Precision reflects the proportion of predicted scam cases that are indeed scams, highlighting the model's ability to avoid false positives. Recall, in contrast, measures the fraction of true scam cases that are correctly detected, indicating how well the model avoids missed detections. Both metrics are critical for practical deployment: high recall ensures effective identification of scams, while high precision reduces disruption for legitimate users.

\begin{figure}[t]
    \centering
    \includegraphics[width=1\linewidth]{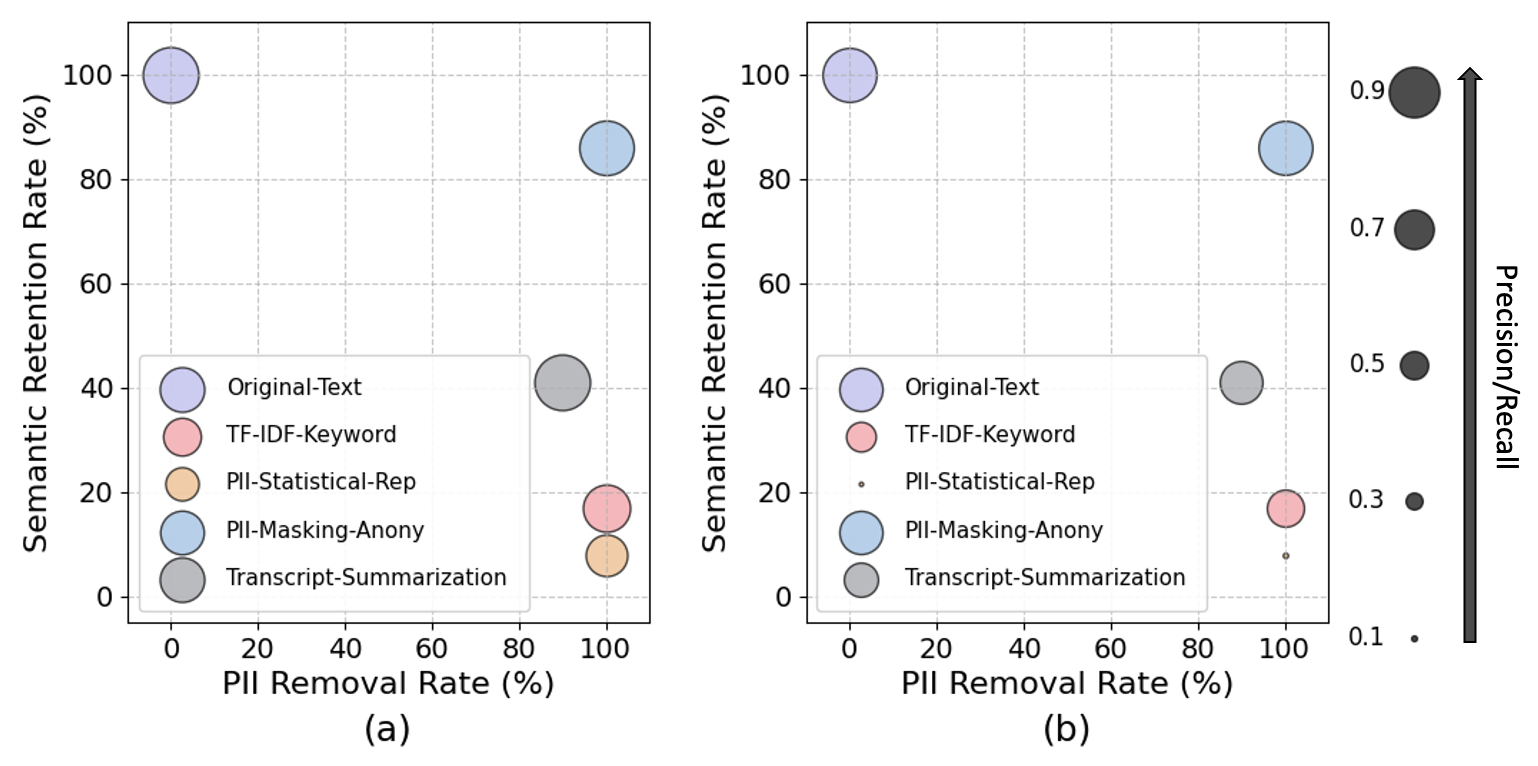}
    \caption{Trade-off between PII removal rate (PRR), semantic retention rate (SRR), and detection performance. (a) Circle size represents precision. (b) Circle size represents recall. Colors indicate different sanitization strategies.}
    \label{fig:tradeoff_prr_srr_precision_recall_labeled}
\end{figure}

Figure~\ref{fig:tradeoff_prr_srr_precision_recall_labeled} visualizes the trade-off between privacy protection and utility for all sanitization methods. Each subplot displays the relationship between PII removal rate (PRR), semantic retention rate (SRR), and detection performance. In panel (a), circle size represents precision; in panel (b), circle size reflects recall. Colors are used to distinguish different sanitization approaches.

The results indicate that the \textbf{Original Text} and \textbf{PII Masking Anonymization} strategies both achieved high precision and recall, clustering in the upper part of each plot with large circle sizes. This suggests that entity-level masking preserves most detection capability while eliminating explicit identifiers. In contrast, both \textbf{TF-IDF Keyword Representation} and \textbf{PII Statistical Representation} achieve high PII removal but exhibit a substantial loss of semantic retention, which is reflected in much smaller circle sizes, especially for precision, indicating lower detection reliability. These methods tend to generate more false positives, largely due to the absence of conversational detail and context needed for robust discrimination between scam and non-scam transcripts.

The \textbf{Transcript Summarization} method occupies an intermediate position, with moderate semantic retention and detection performance. While summarization reduces input redundancy and enhances privacy, the moderate circle sizes indicate some loss of information needed for complete scam detection. The method is suitable in settings where privacy or input length constraints are strict, but users should be aware of the accompanying decline in recall and precision.

Overall, the visual analysis in Figure~\ref{fig:tradeoff_prr_srr_precision_recall_labeled} underscores the need to maintain a balance between privacy protection and detection accuracy. Approaches that retain conversational structure, such as entity-level masking, offer the most reliable compromise. Excessive abstraction or compression, however, tends to compromise one or both critical aspects of detection.

\subsection{Qualitative Comparison of LLM Responses Across Sanitization Methods}

We further investigate how different sanitization strategies affect the reasoning and explanation provided by LLMs in scam detection. Representative model outputs were systematically reviewed to identify how varying levels of abstraction and information loss influence both the accuracy and interpretability of LLM-generated responses.

\textbf{TF-IDF Keyword Representation.}
When the input is represented by TF-IDF keywords, the model primarily focuses on the frequency and presence of specific terms associated with scam scenarios. The explanations tend to be formulaic, with decisions based on whether high-risk words such as "police", "identity card", or "account" are present. This method can sometimes identify scams that match expected vocabulary, but it lacks the ability to interpret context or conversational nuance. As a result, it often produces explanations that are repetitive and shallow, and it can miss less typical scams or produce false alarms when benign conversations contain similar keywords.

\textbf{PII Statistical Representation.}
Using statistical features of personally identifiable information (PII) types, the model's judgments are mainly based on the diversity and frequency of sensitive entities, such as names, numbers, or time references. The LLM typically justifies its output by noting the absence of multiple or high-risk identifiers, for example, the lack of ID numbers, phone numbers, or bank card information. This method is effective for scams that depend on explicit information exchange, but it is limited in its ability to detect scenarios where scams rely more on manipulation or indirect cues. As a result, the explanations are narrow and sometimes overlook relevant behavioral patterns.

\textbf{PII Masking Anonymization.}
When input transcripts are anonymized by replacing sensitive details with category placeholders, the LLM has access to the complete conversational structure minus explicit identifiers. In this setting, model explanations are more detailed and grounded in the sequence and nature of the interaction. The responses often discuss the logic of the conversation, including impersonation, urgency, or attempts to create psychological pressure. This approach enables the model to reference the development of the conversation, escalation, and behavioral indicators, supporting a more reliable and transparent rationale for its predictions.

\textbf{Transcript Summarization.}
Summarized transcripts provide the model with a condensed version of the original conversation. The LLM's responses under this condition are typically brief, highlighting general risk factors such as requests for money or mention of urgency. However, these summaries frequently exclude the subtle cues and interactional context that are sometimes necessary to distinguish complex scams from legitimate disputes. The explanations are less specific, and the detection of nuanced or atypical scams becomes more difficult.

\section{Discussion}

\begin{figure}[t]
    \centering
    \includegraphics[width=1\linewidth]{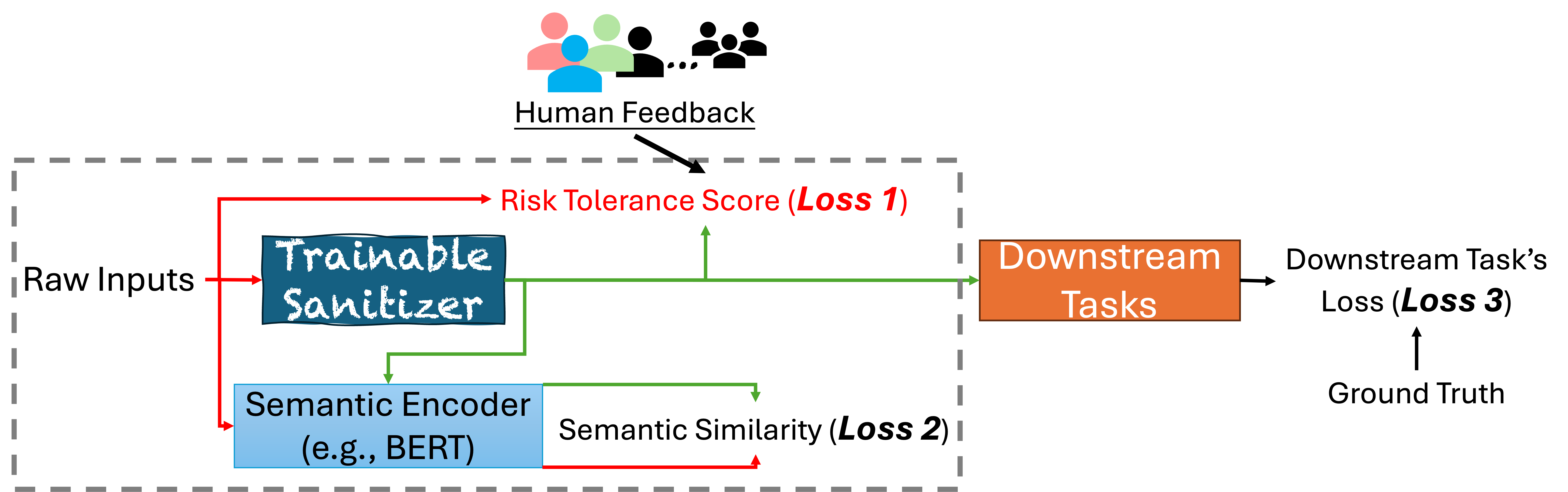}
    \caption{Trainable sanitization framework that jointly optimizes privacy, semantic retention, and task accuracy.}
    \label{fig:trainable_framework}
\end{figure}

\begin{table}[t]
  \centering
  \resizebox{\linewidth}{!}{%
  \begin{tabular}{lcccc}
    \toprule
    \textbf{Local LLM} & \textbf{Accuracy} & \textbf{Precision} & \textbf{Recall} & \textbf{F1 Score} \\
    \midrule
    Gemma3-4b & 0.62 & 0.57 & 0.98 & 0.72 \\
    Qwen3-4b-Think & 0.79 & 0.70 & 1.00 & 0.83 \\
    Qwen3-4b-NoThink & 0.70 & 0.83 & 0.50 & 0.62 \\
    % \midrule
    % \textbf{Cloud LLM} & \textbf{Accuracy} & \textbf{Precision} & \textbf{Recall} & \textbf{F1 Score} \\
    % \midrule
    % ChatGPT-4o & 0.97 & 1.00 & 0.94 & 0.97 \\
    % Gemini-2.5-Flash & 0.96 & 0.96 & 0.96 & 0.96 \\
    % Qwen2.5-Max & 0.98 & 1.00 & 0.96 & 0.98 \\
    % DeepSeek-V3 & 0.98 & 1.00 & 0.96 & 0.98 \\
    \bottomrule
  \end{tabular}%
  }
  \caption{Performances of local LLMs for phone scam detection on original transcripts.}
  \label{tab:exp_results_raw_data}
\end{table}

\subsection{Privacy Preferences and Trainable Sanitization}

Our analysis highlights that the choice of sanitization strategy plays a critical role in balancing privacy protection with the preservation of semantic information essential for accurate downstream detection. 
This insight also suggests that users’ privacy preferences, particularly their individual risk tolerance, are intrinsically linked to detection outcomes. 
Since privacy is inherently subjective and context-dependent, users differ in what information they are comfortable disclosing and how much potential exposure they are willing to accept.
The analysis above demonstrates that MASK can provide a configurable framework that allows users to select sanitization strategies tailored to their personal privacy expectations. 
These strategies operate within a predefined spectrum of privacy-utility trade-offs (e.g., Figure~\ref{fig:tradeoff_prr_srr_precision_recall_labeled}), enabling users to make informed decisions that align with their sensitivity thresholds while maintaining detection effectiveness. 

Further, as depicted in Figure 5, a trainable sanitization module can further enhance MASK's adaptive capability. 
By jointly training the sanitization module with a semantic encoder and a downstream detection model, the system can dynamically optimize for multiple objectives, including privacy risk minimization, semantic fidelity, and detection performance. This multi-objective optimization enables more nuanced control over the trade-off curve between privacy and utility.

\subsection{On-Device Detection with Lightweight LLMs}

While the MASK's ``sanitize-then-detect" paradigm offers a modular and privacy-preserving approach, a natural and naive solution would be to perform scam detection directly on the user's device using a compact LLM. 
This method would eliminate the need to transmit any user data to external servers, thus maximizing privacy by design.

However, as shown in Table~\ref{tab:exp_results_raw_data}, our experiments reveal that running a local lightweight LLM for detection remains challenging in practice. Current small models often lack the necessary accuracy and contextual understanding compared to larger, cloud-based models. Deploying them effectively for real-time phone scam detection, which often involves long, complex conversations, requires significant advances in several areas.

One possible direction is to fine-tune or distill larger models into smaller ones optimized for on-device usage.
However, this approach requires access to large-scale, high-quality labeled datasets, which are currently scarce in the domain of phone scam detection. Moreover, the real-time, streaming nature of phone conversations adds further constraints on latency and memory usage, making the deployment of even moderately sized models challenging on consumer-grade mobile devices.

\section{Conclusion}

This work proposes a novel adjustable sanitization framework that incorporates human preferences in selecting sanitization strategies for privacy-preserving phone scam detection using cloud-based large language models. We explore common sanitization strategies and the results demonstrate an inherent trade-off between privacy protection and the preservation of semantic information required for accurate detection. The framework can be further extended with adaptive, trainable sanitizers to better balance privacy and utility. These findings offer practical guidance for deploying privacy-conscious language understanding systems in sensitive real-world applications.

% \zyq{To be done}
\section{Acknowledgements}
This work was supported the Hong Kong Research Grant Council under Grant 15600219.

\bibliographystyle{ACM-Reference-Format}
\bibliography{main}

\end{document}